\documentclass{article}

\usepackage{microtype}
\usepackage{graphicx}
\usepackage{subcaption}
\usepackage{booktabs} 

\usepackage[utf8]{inputenc}

\usepackage[normalem]{ulem}
\usepackage{amsmath,amsthm,amssymb}
\usepackage{algorithm}
\usepackage[noend]{algpseudocode}
\usepackage{verbatim}
\usepackage{adjustbox}
\usepackage{xpatch,amsthm}
\usepackage{mathtools}
\usepackage{comment}

\usepackage{hyperref}

\usepackage[accepted]{sysml2019}

\input{definitions.sty}

\begin{document}

\twocolumn[
\sysmltitle{Train Where the Data is: A Case for Bandwidth Efficient Coded Training}




\begin{sysmlauthorlist}
\sysmlauthor{Zhifeng Lin}{usc}
\sysmlauthor{Krishna Giri Narra}{usc}
\sysmlauthor{Mingchao Yu}{usc}
\sysmlauthor{Salman Avestimehr}{usc}
\sysmlauthor{Murali Annavaram}{usc}
\end{sysmlauthorlist}

\sysmlaffiliation{usc}{Department of Electrical Engineering, University Of Southern California, Los Angeles, California, USA}

\sysmlcorrespondingauthor{Krishna Giri Narra}{narra@usc.edu}

\sysmlkeywords{Availability, Neural Networks, Deep Learning, Coded Computing, Redundancy, Cloud Computing, Machine Learning}
\vskip 0.3in

\begin{abstract}
Training a machine learning model is both compute and data-intensive. Most of the model training is performed on high performance compute nodes and the training data is stored near these nodes for faster training. But there is a growing interest in enabling training near the data. For instance, mobile devices are rich sources of training data. It may not be feasible to consolidate the data from mobile devices into a cloud service, due to bandwidth and data privacy reasons.  Hence distributed training on mobile devices using data available at the device is gaining traction, as these devices are becoming fairly compute capable. Distributed training at these mobile devices is however fraught with challenges. First mobile devices may join or leave the distributed setting, either voluntarily or due to environmental uncertainties, such as lack of power. Tolerating uncertainties is critical to the success of distributed mobile training. One proactive approach to tolerate computational uncertainty is to store data in a coded format and perform training on coded data. Encoding data is a challenging task since traditional erasure codes require multiple devices to exchange their data to create a coded data partition, which places a significant bandwidth constraint. Furthermore coded computing traditionally relied on a central node to encode and distribute data to all the worker nodes, which is not practical in a distributed mobile setting. 

In this paper, we tackle the uncertainty in distributed mobile training using a bandwidth-efficient encoding strategy. We use a Random Linear Network coding (RLNC) which reduces the need to exchange data partitions across all participating mobile devices, while at the same time preserving the property of coded computing to tolerate uncertainties. We implement gradient descent for logistic regression and SVM to evaluate the effectiveness of our mobile training framework. We demonstrate upto \textasciitilde 50\% reduction in total required communication bandwidth compared to MDS coded computation, one of the popular erasure codes. This improvement is achieved without significantly compromising the tolerance to device failures.

\end{abstract}
]

\section{introduction}
Mobile devices are one of the richest source of data for machine learning model training. There is a growing interest to exploit the data present in these mobile devices to enable distributed training. Distributed training in itself has been extensively used for model training~\cite{ speedUpML,LMA16_unify,reisizadehmobarakeh2017coded,polyCodes,dutta2016short,s2c2}. However, distributed training using mobile devices exacerbates some of the existing bottlenecks, and also bring to forefront new challenges. Distributed training, for instance, has to deal with speed variability across nodes (stragglers). The straggler problem is much worse in a mobile setting due to device power constraints. The second challenge is that the bandwidth available in a mobile setting is significantly lower than in a cloud setting. Hence, approaches that require significant data replication and data exchanges are not well suited for mobile training. Finally, in a distributed mobile environment relying on a single mobile node to act as a master to orchestrate computations and communications is not a viable option.

Before we present the need for a mobile centric solution we discuss how distributed training in the cloud tackles the straggler problem.  One reactive approach to tolerating stragglers in the cloud include task replication. For instance, in distributed training frameworks that use Hadoop MapReduce~\cite{hadoop} and Spark~\cite{spark}, there is a master node that keeps track of training progress on each worker node and then relaunches the work of a slow node on another node, with the restriction that the new worker node must have a copy of the training data partition to be processed. Rather than react to a slow node, recently the idea of distributed training on \textit{coded data} has been proposed as a proactive strategy~\cite{speedUpML,LMA16_unify,reisizadehmobarakeh2017coded,polyCodes,dutta2016short, s2c2}.  This strategy is known as coded computing. In coded computing, redundancy is added in an efficient coded form to make the computations robust to stragglers. In ~\cite{speedUpML} error correcting codes such as Maximum-Distance-Separable (MDS) codes are utilized to create redundant computations during gradient descent. For example, a $(N,K)$-MDS coding allows computation to be distributed to $N$ nodes, but can recover the full result from any $K < N$ nodes that complete their computation. Hence, MDS in this case can tolerate \textit{any} $N-K$ nodes to be stragglers.

Coded computing has the benefit that it is proactive and does not wait for a failure to relaunch a task. It does not require a priori knowledge of which node may fail and hence removes the guess work from identifying which data partitions must be replicated. Furthermore, recent innovations in coded computing~\cite{s2c2, seqApp, YaoqingNIPS2017, mallick2019fast} adapt the level of redundant computing to the speed variability thereby reducing the overhead of coded computing. As such coded computing is one viable option for adaption to a mobile training setting.

However, existing coded computing solutions assume a centralized execution model, namely, there is a master node which can access all the data to encode and distribute. In a mobile setting data is inherently collected and stored in distributed manner among mobile devices; and a mobile node acting as a master is often incapable of accomplishing data encoding due to highly constrained compute, memory and storage resources (we will justify this assertion quantitatively in section \ref{motivation}). Second, MDS coding requires that some of the encoded data partitions be built from \textit{all} the distributed data partitions. Thus all the mobile devices must transmit their data to a single node to create the encoded partition, which leads to bandwidth bottlenecks, as we describe in detail in our background section later. 

Based on these observations, in this paper we explore bandwidth efficient distributed training on mobile devices. We use random linear network coding (RLNC) to achieve this goal. Our goal is to quantify the benefits of using different coding strategies in quantifying the bandwidth demands of gathering all data partitions to create an encoded partition, rather than propose new coding algorithms. We demonstrate how RLNC can be used to eliminate the single master bottleneck to distribute the encoding task of the data to a set of mobile devices, while concurrently limiting inter-device communication cost and encoding complexity. However, RLNC comes with a slight degradation in straggler tolerance, compared to MDS, and we alleviate this issue by increasing the computational redundancy by a very small fraction to match the straggler tolerance of MDS coded training. 

In summary, the main contributions of this paper are as follows: 
\begin{itemize}
\item We built a 22-node mobile computing cluster consisting of raspberry pi nodes to perform distributed training of support vector machine and linear regression models, models that are better suited for mobile compute limitations. We provide an analysis of the computation and communication limitations which make centralizing encoding and distribution infeasible in a mobile computing cluster.  

\item We show the drawbacks of MDS based distributed encoding and propose RLNC coded computation, which leverages the presence of training data within a mobile node and provides distributed and local encoding of data. Its code structure enables reductions in both the amount of communication needed for encoding and also the encoding complexity.

\item We evaluate the performance of RLNC coded distributed training on our mobile cluster. We demonstrate that RLNC outperforms the conventional MDS-coded computation. RLNC achieves \textasciitilde 50\% reduction in total required communication bandwidth compared to MDS coded computation. RLNC reduces the encoding latency by \textasciitilde 50\% compared to MDS coded computation. These improvements are achieved without significantly compromising on straggler tolerance.
\end{itemize}

The organization of the rest of the paper is as follows: In section 2 we briefly define coded computing used in model training. In section 3 we demonstrate the difficulties of implementing centralized encoding in a distributed mobile environment. In Section 4, we introduce our RLNC coded distributed training framework. In section 5 we provide implementation and system details. In section 6 we provide evaluation and results. In Section 7 we describe the related work. We conclude the paper in Section 8.

\section{Coded Distributed Training}\label{background}
\subsection{Brief Primer on Erasure Codes}\label{subsec:erasureCode}
Erasure codes are a major class of error correcting codes which have been adopted recently for dealing with straggler uncertainties in distributed training. An erasure code generates $N$ linear combinations (a.k.a. coded symbols) from $K$ information symbols, so that when $K$ linearly independent coded symbols are received, the $K$ information symbols can be decoded through solving linear equations. The code can be described by a $K\times N$ \emph{generator matrix} $\mG$:
\begin{equation}
\mG=\left[
\begin{matrix}
\a_{0,0} & \a_{0,1} & \cdots & \a_{0,N-1}\\
\a_{1,0} & \a_{1,1} & \cdots & \a_{1,N-1}\\
\vdots & \vdots & \ddots & \vdots\\
\a_{K-1,0} & \a_{K-1,1} & \cdots & \a_{K-1,N-1}\\
\end{matrix}
\right],
\end{equation}
where $\{\a_{k,n}\}$ are scalars, and column-$n$ is the coefficient vector for the $n$-th coded symbol. If every set of $k$ columns are linearly independent, then $\mG$ is called an MDS (maximum distance separable) matrix, which allows the recovery of $K$ information symbols from the value of any $K$ out of $N$ coded symbols. This property of MDS codes, any $K$ out of $N$ are sufficient for recovery, makes it the best code for dealing with straggler servers in distributed computing. An example of MDS codes, under real numbers, is the classic Reed-Solomon code, where  $\{\a_{k,n} = (n+1)^{k}\}$. 

\subsection{Application to Distributed Training}

We can apply MDS codes to encode distributed training for models such as SVM and linear regression. Matrix algebra is the foundation for training these models. Note that coded computing for deep neural network (DNN) training is an ongoing work in academia. For a large matrix-vector multiplication $\mA \cdot \x$, provided $\mG$ in a $(N,K)$-MDS code,  $\mA$ is partitioned into $K$ equal-height sub-matrices $\mA_0,\cdots$, $\mA_{K-1}$. Each worker $w_n$ ($n\in[0,N-1]$) computes $\mtA_n\cdot \x$, where $\mtA_n=\sum_{k=0}^{K-1}\a_{k,n}\mA_k$ according to the associated column $\mG[:,n]$. Whenever the fastest $K$ workers have completed (since MDS ensures any $K$ columns of $\mG$ are linearly independent), we can use their results to decode the value of $\mA_0\cdot\x,\cdots,\mA_{K-1}\cdot\x$, and then recover the value of $\mA\cdot\x$ through vertical concatenation.

\begin{figure}
\centering
\includegraphics[trim ={1.75in .25in 1.70in 1.75in}, clip,width=.5\textwidth, height=7cm]{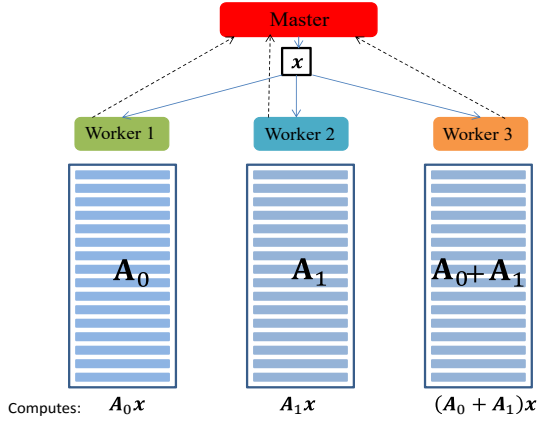}
\caption{An illustration of CDC for $\mA\cdot \x$.}
\label{cdc}
\end{figure}

Figure \ref{cdc} shows an example using the MDS code for computing $\mA\cdot \x$ on three machines with one potential straggler; $\mA$ is a large input matrix and $\x$ is a vector. Matrix $\mA$ is horizontally partitioned into two equal-height sub-matrices $\mA_0$ and $\mA_1$. Worker $w_0$ computes $\mA_0\cdot \x$. Worker $w_1$ computes $\mA_1\cdot \x$. Worker $w_2$ computes $\mtA_2\cdot \x$, where $\mtA_2=\mA_0+\mA_1$ is from encoding. The completions of any $2$ out of the $3$ workers allow the recovery of $\mA_0\cdot \x$ and $\mA_1\cdot \x$ and, thus, $\mA\cdot \x$.

This figure corresponds to the following $(N, K)$-MDS generator matrix $\mG$ with $K=2$ and $N=3$. $\mA_0$ is encoded using the first column of $\mG$ and consists of top half rows of $\mA$. $\mA_1$ corresponds to the second column of $\mG$ and consists of lower half rows of $\mA$. $\mtA_2 = \mA_0+\mA_1$ corresponds to the third column of $\mG$.
\begin{equation*}
\mG=\left[
\begin{matrix}
1 & 0 & 1\\
0 & 1 & 1\\
\end{matrix}
\right].
\end{equation*}

Another $(N, K)$-MDS generator matrix $\mG$ with $K=2$ and $N=4$ shown below. The first 3 sub-matrices are identical to the ones in previous example. The sub-matrix corresponding to the last column of $\mG$ is $\mtA_3 = \mA_0+2\mA_1$. Here when any two workers out of the four complete their computations the complete result can be decoded. 

\begin{equation*}
\mG=\left[
\begin{matrix}
1 & 0 & 1 & 1\\
0 & 1 & 1 & 2\\
\end{matrix}
\right].
\label{eq:GofMDSEx2}
\end{equation*}

\subsection{A Centralized Master-Workers Model}\label{subsec:MW}
When deploying coded computing for distributed training, prior solutions proposed in \cite{LMA16_unify,reisizadehmobarakeh2017coded,polyCodes,dutta2016short} require a Master-Workers model. This model assumes that the master has possession of all the data, or in other words, the entirety of matrix $\mA$. Given a $(N, K)$ code, the master will partition $\mA$ into $K$ sub-matrices, linearly combine these $K$ partitions to create encoded sub-matrices, and then distribute them to the $N$ workers in the system. Upon receiving \emph{any} $K$ results out of the $N$ workers, the master will be able to decode and reconstruct the final value of $\mA \cdot \x$.

In the next section, we provide a quantitative analysis of the cost for using a centralized model in a distributed mobile setting.

\section{Cost of Mobile Training}\label{motivation}

To evaluate the cost of MDS coding in a distributed mobile setting we built a 22-node mobile cluster consisting of raspberry pi nodes. In our experiments, we  consider a single Raspberry Pi 3 Model B processor as representative of a low cost and ubiquitous mobile device. 
We implemented the single master-many workers model described in \ref{subsec:MW} on the mobile cluster to perform coded matrix-vector multiplication and measured few performance metrics. We measured the time it takes to load a matrix partition from into memory in the mobile device and a Xeon server node for two $(N,K)$ coding configurations: (22,12), (22,16). The dimensions of the matrix are 14000 rows $\times$ 5000 columns. We assume that the data partitions to be loaded are communicated to a single master node which stores the data and then loads it to main memory. For a (22,12)-MDS code this matrix is divided into 12 partitions, and for (22,16)-MDS the matrix is divided into 16 partitions.

Table \ref{tab:rpiLoad} shows the measured values for load time (time to read storage and bring just one partition of matrix data to main memory).  It can be observed that the time to load one partition is two orders of magnitude slower on the mobile node when compared to the corresponding time on a server node. 

Next we measured the time to encode a single partition in the master node in the two $(N,K)$ configurations: (22,12), (22,16). Note that the encoding process in MDS  sometimes requires just taking one partition of the matrix and passing it to a worker node as is, and in other cases the encoding process has to compute a linear combination of many sub-matrices to create a single encoded matrix that is then sent to a worker node. The data shows the simplest of such a linear encoding task that performs the encoding operation $\mtA=\mA_0+\mA_1+..+\mA_{K-1}$. These values are shown in table \ref{tab:rpiEncode}. The mobile node spends time that is again two orders of magnitude more to encode this linear summation when compared to the server node. This result shows that when a mobile node is responsible for doing the encoding it takes significant amount of latency even before a worker node is initiated for computation. Thus relying on a single master node to do encoding and distribution makes it a severe bottleneck in distributed mobile training.

\begin{table}[]
    \centering
    \begin{tabular}{|c|c|c|}
    \hline
    Time to load one partition& Xeon& Raspberry pi\\
    \hline
    (22,12) & 2.8 & 484 \\
    \hline
    (22,16) & 2.15  & 367 \\
    \hline
    \end{tabular}
    \caption{Load Time in seconds}
    \label{tab:rpiLoad}
\end{table}
\begin{table}[]
    \centering
    \begin{tabular}{|c|c|c|}
    \hline
    Time encoding one block& Xeon& Raspberry pi\\
    \hline
    (22,12) & 43.29 & 3358 \\
    \hline
    (22,16) & 43.74 & 3300 \\
    \hline
    \end{tabular}
    \caption{Encode Time in seconds}
    \label{tab:rpiEncode}
\end{table}

Also in a mobile device cluster, it is unlikely that data to be processed is available centrally in a single node. Each data partition is collected and stored at its corresponding mobile node. To apply the master-workers computing model in this setting, all the data from mobile nodes needs to be sent to a single master node first before encoding. This would lead to huge increase in communication and is impractical in a mobile setting.

Due to these reasons we consider alternate distributed computing strategies. One straight forward strategy is to \emph{not} do encoding at the master node and let the mobile nodes communicate data among themselves and perform encoding locally. In this distributed encoding strategy each mobile node uses the same \emph{generator matrix}. The mobile nodes then communicate, exchange data among themselves and locally generate their encoding matrices. Each node $i$ uses column $i$ of matrix $\mG$ to determine which partitions are needed to encode its data. For instance, if  column $1$ has $1,0,0..0$ then the mobile node $1$ needs only the first partition. On the other hand, if column $i$ has $1,2,3..K$ then the worker $i$ needs to get all partitions and then multiply 2nd partition by 2, 3rd partition by 3, and so on and then sum all to get the encoded matrix. 

There are several approaches to create a generator matrix for MDS. Most of them would need all $N$ mobile nodes to download few partitions from other nodes. If we apply systematic MDS coding as discussed in prior work \cite{speedUpML}, then $\mG$  has $K\times K$ identity matrix as the first $K$ columns of its $\mG$, followed by $N-K$ columns where each element in the column is a non-zero value . For example, the systematic MDS applied in our implementation is of the form:
\begin{equation}
\mG=\left[
\begin{array}{cccc|ccccc}
1 & 0 & \cdots & 0 & 1 & 1 & 1 & \cdots\\
0 & 1 & \cdots & 0 & 1 & 2 & 3 & \cdots\\
0 & 0 & 1 & \ddots & 1 & 3 & 5 & \cdots\\
\vdots & \vdots &\ddots & \vdots & \vdots & \vdots \\
0 & 0 & \cdots & 1 & 1 & K & 2K-1 & \cdots
\end{array}
\right],
\end{equation}
Due to the identity matrix, systematic MDS code relieves the need of the first $K$ mobile nodes for encoding. They simply have to select the partition that they already have. However, all the remaining entries of $\mG$ must be nonzero to maintain MDS, indicating that all the remaining $N-K$ mobile nodes must download and process \textit{all $K$ sub-matrices} to generate their own encoded sub-matrices. This communication process is shown visually in figure \ref{codedExample} (left hand figure). In this figure there are six data partitions each located at one of the six nodes, and we use (8,6)-MDS code. Hence the two nodes must access all the six partitions to encode the matrix, which is a significant cost. In general applying $(N,K)$ MDS code as described in prior work \cite{speedUpML} enables  optimal recovery from stragglers, namely decoding after any $K$ completed workers as explained in section \ref{subsec:erasureCode}. However, the price is high computation and communication costs due to a dense generator matrix.

Consequently, if we rely on MDS codes to perform distributed training on mobile devices we 
suffer from:
\begin{itemize}
    \item high communication cost, because some mobile nodes must download \textit{all} the sub-matrices. 
    \item high computation cost, because 1) some mobile nodes must linearly combine all the sub-matrices; and 2) the coefficients could be prohibitively large.
\end{itemize}
While these drawbacks are present even in the distributed training in the cloud, prior approaches have used MDS (and its derivatives) extensively because they assume that all data partitions are available to the master and the master is powerful enough to quickly encode the matrices. However, in  mobile training environment data is dispersed and communication resources are limited. In this paper, we propose to evaluate systematic binary random linear network coding (RLNC) \cite{heide_systematic_RLNC} as a more viable coding option in a mobile environment. We describe this coding in the next section.

\section{Mobile-Centric RLNC Coding}
Given $(N,K)$, the construction of generator matrix $\mG$ is much simpler for RLNC: The first $K$ columns of $\mG$ are set as an identity matrix, similar to systematic MDS, and the entries of the remaining $N-K$ columns are randomly set to 0 or 1 with a equal probability of 50\%. For example, a randomly generated instance of $\mG$ with $K=6$ and $N=8$ is:
\begin{equation}
    \mG=\left[
    \begin{array}{cccccc|cc}
    1 & 0 & 0 & 0 & 0 & 0 & 1 & 0 \\
    0 & 1 & 0 & 0 & 0 & 0 & 0 & 1 \\
    0 & 0 & 1 & 0 & 0 & 0 & 0 & 1 \\
    0 & 0 & 0 & 1 & 0 & 0 & 1 & 0 \\
    0 & 0 & 0 & 0 & 1 & 0 & 1 & 0 \\    
    0 & 0 & 0 & 0 & 0 & 1 & 0 & 1
    \end{array}
    \right]
\end{equation}
Under RLNC, the first $K$ mobile nodes do not need any download or coding (similar to MDS), and the remaining $N-K$ mobile nodes only download and encode an average of $\frac{K}{2}$ sub-matrices. Thus, RLNC can significantly reduce both the communication bandwidth and computation costs by 50\% on average. Each mobile node that needs to encode the data  simply sums the sub-matrices whose coefficient is 1. Since no large coefficients are involved, RLNC is more appealing for mobile devices.

\begin{figure}
\centering
\includegraphics[width=.5\textwidth, height=6.5cm]{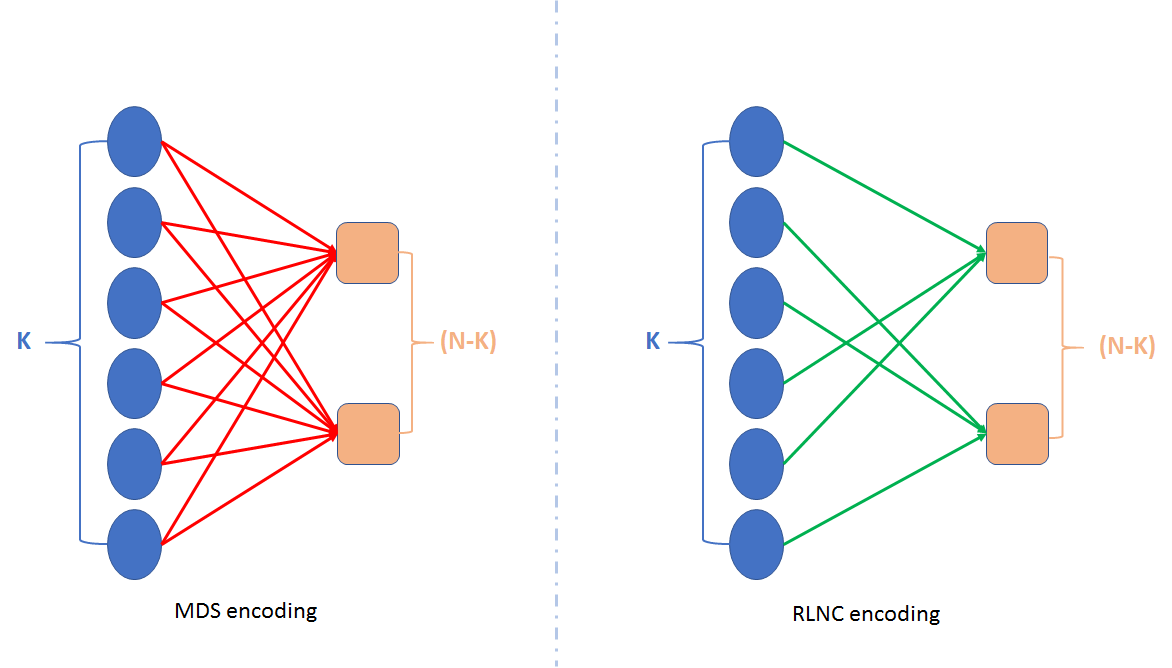}
\caption{An illustration of encoding patterns of MDS v.s. RLNC. MDS requires all k sub-matrices; whereas on average, RLNC requires only $\frac{k}{2}$ sub-matrices}
\label{codedExample}
\end{figure}

In figure \ref{codedExample} (right figure), we draw the download dependencies with RLNC. From this figure, we can observe that RLNC reduces 50\% of the communication bandwidth since any mobile device only needs half of the data partitions from other devices.

\begin{figure}
\centering
\includegraphics[width=.5\textwidth, height=5.5cm]{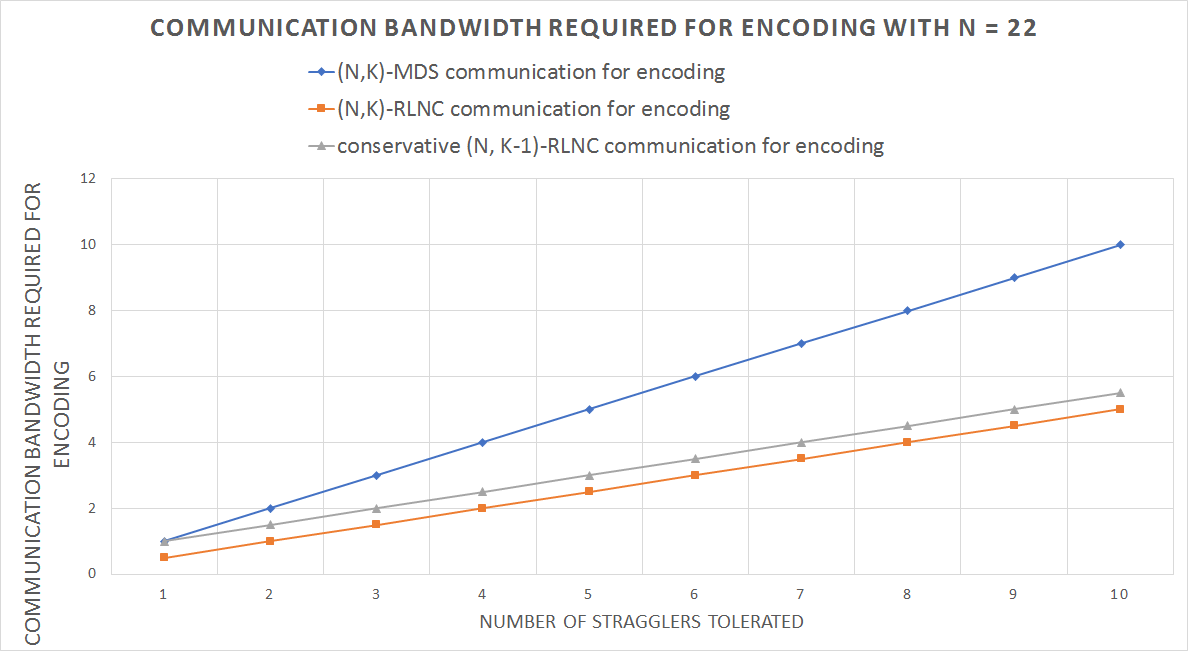}
\caption{Communication Bandwidth Used for MDS, RLNC and a more conservative RLNC}
\label{fig:conservative}
\end{figure}

In figure \ref{fig:conservative}, we plot the bandwidth usage during distributed encoding while using the MDS and RLNC coding in our mobile cluster with 22 nodes. x-axis in the figure shows the number of stragglers tolerated i.e., the $(N-K)$ value. y-axis in the figure shows the amount of data exchanged, in terms of the number of data partitions, at each coding configuration after being normalized to the size of the matrix. As  observed, RLNC coding reduces the amount of data communicated by half in comparison to MDS coding in all the coding configurations.

\begin{figure}
\centering
\includegraphics[width=.45\textwidth, height=5.5cm]{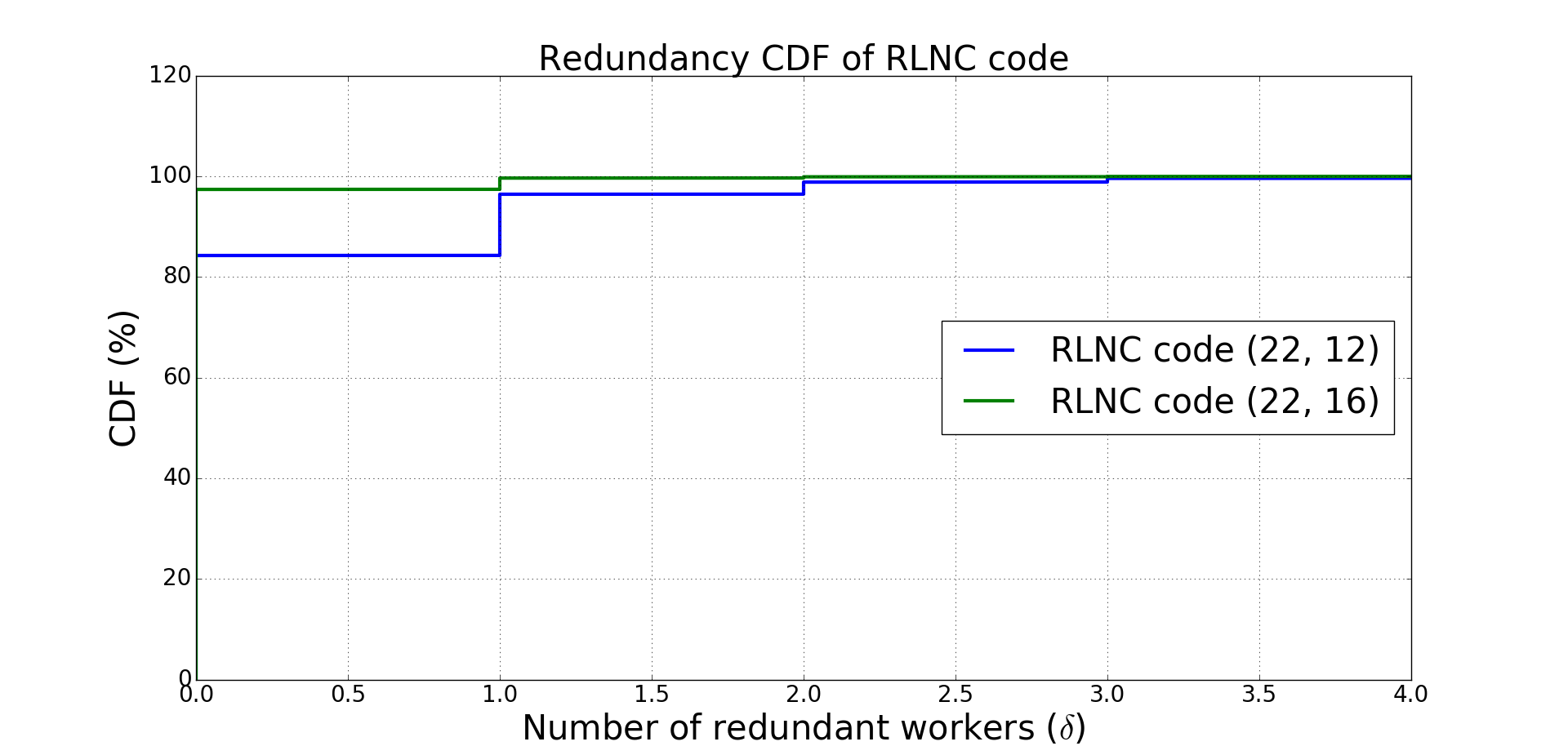}
\caption{Empirical CDF for (22,12) and (22,16) RLNC codes}
\label{cdf}
\end{figure}

On the other hand, due to the random coding, RLNC cannot guarantee decoding after any $K$ completed workers. RLNC allows decoding after any $K+\delta$ completed workers. We ran monte carlo simulations to find the distributions for two RLNC coding schemes, $(22, 12)$ and $(22, 16)$. Their empirical cumulative probability distributions are shown in figure \ref{cdf}. With more than $95\%$ probability, $\delta = 1$, or in other words, only one additional worker will be required to completely decode the result with RLNC coding. And on average, $\delta = 0.2132$ for $(22,12)$-RLNC codes, and $\delta = 0.032$ for $(22,16)$-RLNC codes. This overhead is negligible for any practical value of $K$, and thus makes RLNC asymptotically optimal in terms of recovery while enabling low cost encoding to enable mobile training.

Next we evaluate a more conservative RLNC code, to compensate the fact that RLNC may require $\delta=1$ additional redundant workers to decode with more than $95\%$ probability. We use a $(N, K-1)$-RLNC code instead of $(N,K)$-RLNC code to provide the same level of straggler tolerance as $(N,K)$-MDS code. Even in this scenario, the communication bandwidth requirement is still close to 50\% of what a $(N,K)$-MDS code would require, when a large number of stragglers need to be tolerated (i.e. $N-K$ is large), which is sensible in the highly uncertain mobile computing setting. This is because the ratio of the communication bandwidth required between $(N,K)$-MDS and $(N, K-1)$-RLNC is $\frac{N-K+1}{2(N-K)}$, which can be simplified to $\frac{1}{2} + \frac{1}{2(N-K)}$. In figure~\ref{fig:conservative}, we show the communication bandwidth used for $(N,K)$-MDS, $(N,K)$-RLNC and a more conservative $(N,K-1)$-RLNC code with $N=22$ under different straggler tolerance levels. We can observe that the conservative $(N,K-1)$-RLNC code is also able to reduce the amount of data to be communicated when compared to $(N,K)$-MDS, and when $N-K$ is large, the communication bandwidth required is close to 50\% of what $(N,K)$-MDS would require. 

\noindent \textbf{Fallback Mechanism}: In the worst case scenario, RLNC may need results from more than 1 additional redundant worker to do recovery i.e., decode the final product. For example, as shown in figure~\ref{cdf}, $(22,16)$-RLNC code may need results from 4 additional redundant workers to recover with 100\% probability. If these redundant workers also become straggles in a rare case, instead of waiting on the stragglers, the tasks assigned to these stragglers can be replicated as has been done in Hadoop and Mapreduce frameworks. This fallback mechanism would guarantee forward progress with 100\% probability.

\section{Implementation}
In this section we briefly describe the applications that we evaluated. 

\subsection{Computing Applications}
Gradient descent is popularly used for training machine learning and deep learning models. It is an iterative optimization algorithm which uses matrix-vector multiplication during each iteration. So we evaluated our proposed \emph{mobile training framework} on two machine learning algorithms: Logistic Regression and Support Vector Machine (SVM). Next we describe their implementations in our system.
\subsubsection{Logistic Regression}\label{LRSection}
Logistic regression is a machine learning algorithm widely used to do binary classification. It uses a sigmoid function $ \sigma(a) = \frac{1}{1+e^{(-a)}}$ to classify.
In general, gradient descent has the following form: if \(\boldsymbol{w}\) is the parameter of interest, \(\eta\) is the learning rate, \(\lambda\) is the regularization coefficient, \(\boldsymbol{t}\) is the iteration variable: \(\boldsymbol{w}^{(t+1)} = \boldsymbol{w}^{(t)} - \eta (\nabla \boldsymbol{w}^{(t)} + \lambda\boldsymbol{w})\). 
For logistic regression where we have \(\boldsymbol{X}\) as the matrix representing all the training features and \(\boldsymbol{y}\) as the vector representing training class labels, the gradient can be written as: \(\nabla \boldsymbol{w}^{(t)} = \boldsymbol{X}^{T} 
(\sigma(\boldsymbol{X}\boldsymbol{w}^{(t)})-\boldsymbol{y})\). 

We can observe that there are two major matrix-vector multiplications, \(\boldsymbol{X}\boldsymbol{w}^{(t)}\) and \(\boldsymbol{X}^{T} \boldsymbol{p}\) where \(\boldsymbol{p} = \sigma(\boldsymbol{X}\boldsymbol{w}^{(t)})-\boldsymbol{y}\).
The gradient descent procedure implemented in our system for Logistic Regression is described in Algorithm \ref{alg:LR}. 
During each iteration of Gradient Descent these two matrix-vector products are used to calculate the gradient of the weight vector. In our framework, the two matrix-vector multiplications are coded, using RLNC coding, and parallelized by computing partial matrix-vector products at different workers and then constructing the total matrix-vector product at the end of each iteration.

\begin{algorithm}
\caption{Gradient Descent for Logistic Regression}
\label{alg:LR}
\begin{algorithmic}
\State \textbf{Input:} Dataset $D(i)$ for $i = 1$ to $K$ that are stored distributedly in $K$ workers, $(N,K)$ Coding Strategy $Coding$, Number of iterations $numIter$
\State \textbf{Output:} weight vector of the trained model $w$
\State Preprocess the dataset $D(i)$ and generate sub matrix $X(i)$, $X^{T}(i)$ for $i = 1$ to $K$ and class labels $y$
\State Encode $X(j)$, $X^{T}(j)$ as per $Coding$ for $j = K+1$ to $N$ distributedly at the remaining $(N-K)$ workers 
\State Initialize vector $w$ to random values
\For {$iter$ = 0 to $numIter$}
    \State compute \(\boldsymbol{X}\boldsymbol{w}^{(t)}\) by performing coded computation 
    \State p = $\sigma(\boldsymbol{X}\boldsymbol{w}^{(t)})-\boldsymbol{y}$
    \State compute $\nabla \boldsymbol{w}^{(t)} = \boldsymbol{X}^{T} \boldsymbol{p}$ by performing coded computation 
    \State  $\boldsymbol{w}^{(t+1)} = \boldsymbol{w}^{(t)} - \eta (\nabla \boldsymbol{w}^{(t)} + \lambda\boldsymbol{w})$
    \State $iter \leftarrow iter + 1$
\EndFor
\State return $w$
\end{algorithmic}
\end{algorithm}


\subsubsection{SVM}
Support Vector Machine (SVM) is another commonly used supervised learning model for binary classification of data. For SVM where we have \(\boldsymbol{X}\) as the matrix representing training features and \(\boldsymbol{y}\) as the vector representing training labels, the model can be represented as: \(\boldsymbol{s} = \boldsymbol{X}\boldsymbol{w}^{(t)}\). By considering N as the total number of samples in training data, the sub-gradient can be expressed as: \(\nabla \boldsymbol{w}^{(t)} = \frac{1}{N}\boldsymbol{X}^{T} \boldsymbol{m}\), where m is a vector with \(\boldsymbol{m}_j = -\boldsymbol{y}_j \text{ if \(\boldsymbol{y}_j\boldsymbol{s}_j < 1\) otherwise 0}\).
We can observe that there are two major matrix-vector multiplications here, \(\boldsymbol{X}\boldsymbol{w}^{(t)}\) and \(\boldsymbol{X}^{T} \boldsymbol{m}\). These matrix-vector multiplications are parallelized by computing partial matrix-vector products at different workers and then constructing the total matrix-vector product. 


\subsection{System Setup}
\begin{figure}
    \centering
    \includegraphics[width=\linewidth]{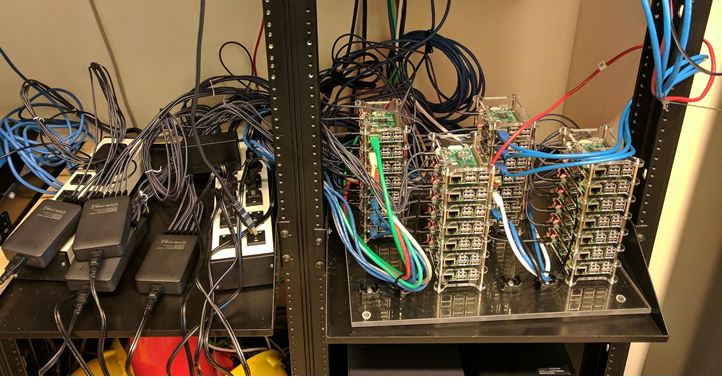}
    \caption{Picture of our Edge Cluster}
    \label{fig:rpis}
\end{figure}

\begin{figure}
    \centering
    \includegraphics[width=\linewidth, height=6cm]{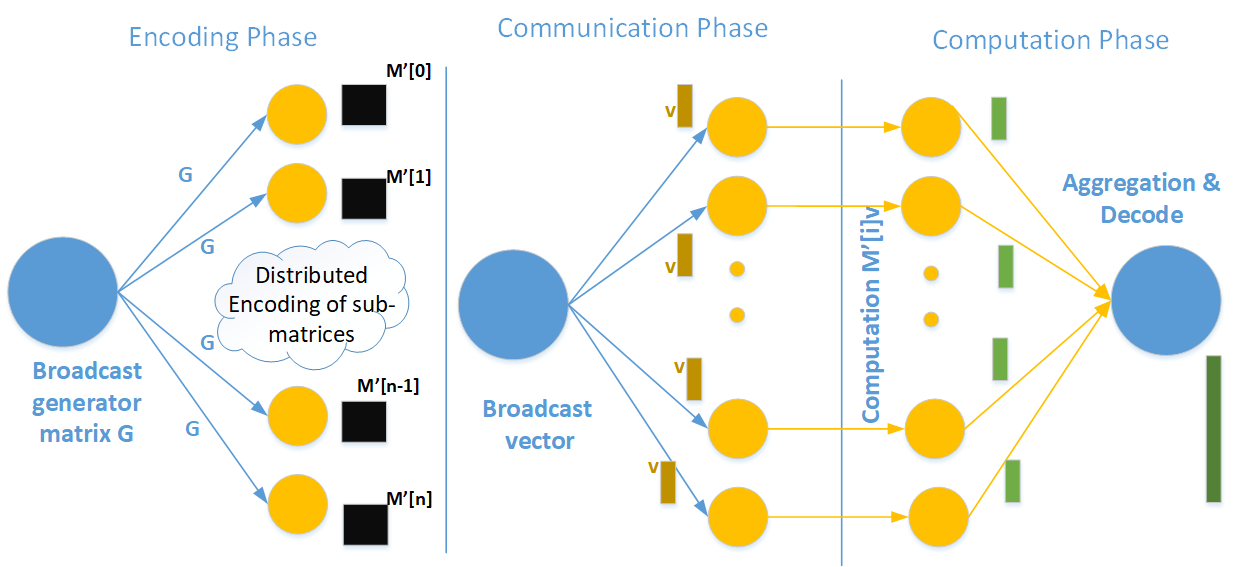}
    \caption{Phases in Distributed Mobile Training}
    \label{fig:phases}
\end{figure}

We evaluated the above computing applications on a Raspberry Pi cluster, which mimics a representative edge computing environment. This cluster is composed of 22 Raspberry Pi 3 Model B nodes. Each node consists of a Quad Core 1.2GHz Broadcom BCM2837 64bit CPU, running HypriotOS V1.7.0 with Linux Kernel 4.4.50. Each Raspberry Pi has 1GB of DRAM and a 64G Western Digital PiDrive connected through USB. And all of the 22 nodes are connected under a wireless network. We bootstrap this Raspberry Pi cluster with Kubernetes and we dockerize our computing applications to run on the Kubernetes cluster. Figure \ref{fig:rpis} is a picture of our Raspberry Pi cluster system. 

\subsection{Implementation details}
Now we describe the workings of our coded computation framework, which is shown in figure \ref{fig:phases}. 22 docker containers are launched onto the Raspberry Pi cluster, representing the workers of the computation task. Each of these worker docker containers is running on a separate Raspberry Pi node. Another docker container is launched within the same cluster to represent the master. Unlike in~\cite{speedUpML} where the master needs to encode and distribute the coded sub-matrices to the workers, our master docker container is simply responsible for coordinating the workers and decoding the matrix-vector multiplication product at the end of each iteration. Since the master docker container is a lightweight coordinator and decoder, it can easily coexist with any of the worker docker containers on the same Raspberry Pi node. 

At the beginning of computation, master sends the $(N, K)$ generator matrix $G$ to each of the workers. While the first $K$ workers have original sub-matrix and do not require encoding, the remaining $(N-K)$ workers request and download their needed sub-matrices from the first $K$ workers to perform encoding based on the generator matrix $G$. Due to the highly constrained memory space, the remaining $(N-K)$ workers need to load the sub-matrices sequentially to do the required encoding. 

\begin{algorithm}[H]
\caption{One iteration for coded training from a master}
\label{alg:codedComputing}
\begin{algorithmic}
\State Master broadcast $V$ to all workers at the beginning of an iteration
\State $set \leftarrow \phi$
\State $partialP_{dict} \leftarrow \phi$
\While {$set$ is not decodable}
    \State \textbf{on} Receiving completion of partial product $p$ computation message from worker $w$
    \State $set \leftarrow (set, w)$
    \State $partialP_{dict} \leftarrow (partialP_{dict}, p_{w})$
\EndWhile
\State Cancel computation at the remaining stragglers
\State $P \leftarrow decode(set, partialP_{dict})$
\end{algorithmic}
\end{algorithm}

One iteration of our coded training is described in \ref{alg:codedComputing}. At the start of each iteration of our applications, the master distributes the vector to be multiplied to all workers and informs the workers to start their computation tasks. At the end of each iteration, master receives the sub-products from the workers, decodes them and constructs the result vector. Each worker has two separate running processes, one for computation and one for communication. The computation process on the worker is straightforward. During each iteration it simply multiplies the sub-matrix with the vector received from the master, generates the partial product and then waits for the next iteration of computation. The communication process is in charge of receiving input data from the master, sending the partial product, and controlling the start and stop of the computation process at the worker.


\section{Evaluation}
Next we discuss the results from our experiments. In our experiments, we emulate stragglers by reducing the performance of a subset of randomly selected nodes. In our Raspberry Pi Cluster, we evaluate the two computing applications described in the previous sections with RLNC coding strategies. For logistic regression, we use a $(22, 16)$-RLNC code to tolerate up to 6 stragglers. And for SVM, we use a more conservative $(22, 12)$-RLNC code, which can tolerate up to 10 stragglers in the 22-node Raspberry Pi cluster.
For both applications, we used the dataset from UC Irvine machine learning repository \cite{uci} and we run Gradient Descent for 100 iterations and repeat the experiments multiple times. We then plot the average of these multiple executions. We use MDS-coded computation as a baseline strategy. We evaluate the required communication bandwidth for data encoding, data loading and encoding time, and the total execution time to finish 100 iterations of Gradient descent for logistic regression and SVM.

\subsection{Communication bandwidth requirement}
First we compare the communication bandwidth requirements of MDS and RLNC for data encoding. As can be seen from figure \ref{fig:conservative}, with $N = 22$, on average RLNC can save $50\%$ of the bandwidth requirement regardless of the number of stragglers we design the code to tolerate when compared to the MDS coding strategy. This is due to its nice construction of the generator matrix $\mG$, where the entries in the the remaining $N-K$ columns are randomly set to 0 or 1 with an equal probability of $50\%$. This translates into the property of our mobile training framework that each of the $N - K$ redundant workers only needs to request and download half of the sub-matrices from the first $K$ workers.

\subsection{Data loading and encoding}
\begin{figure}
\centering
\includegraphics[width=.5\textwidth]{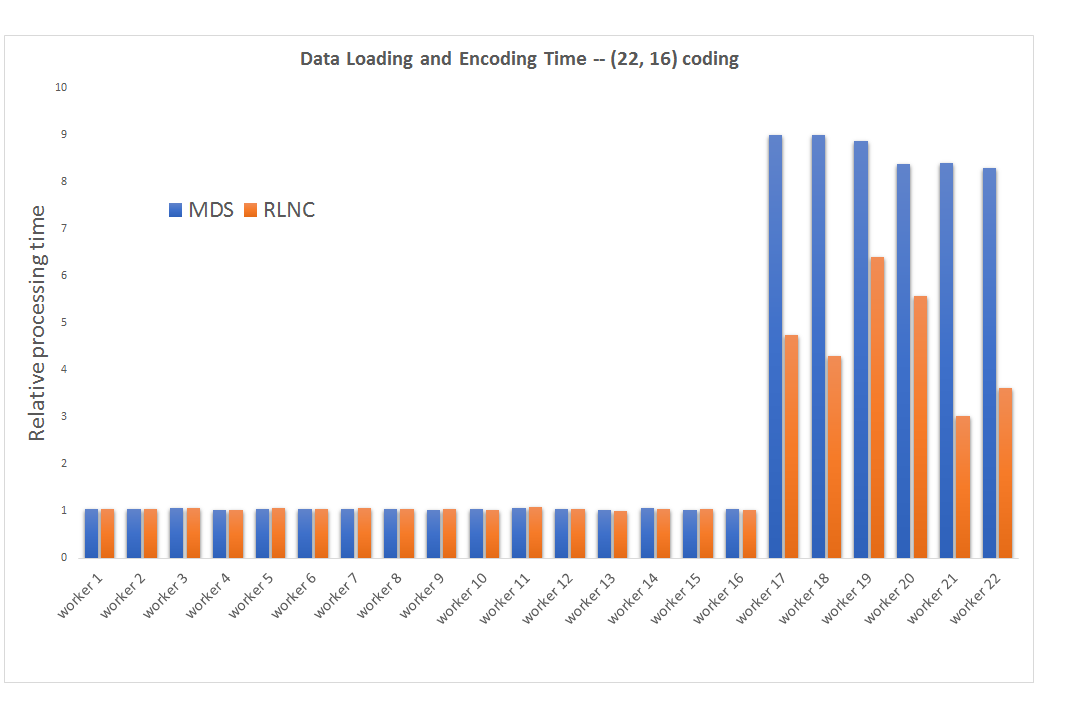}
\caption{Data loading and encoding time on each worker docker container for $(22,16)$ coding. On average, our RLNC can reduce the data loading and encoding time by \textasciitilde 50\%}
\label{encoding_22_16}
\end{figure}

\begin{figure}
\centering
\includegraphics[width=.5\textwidth]{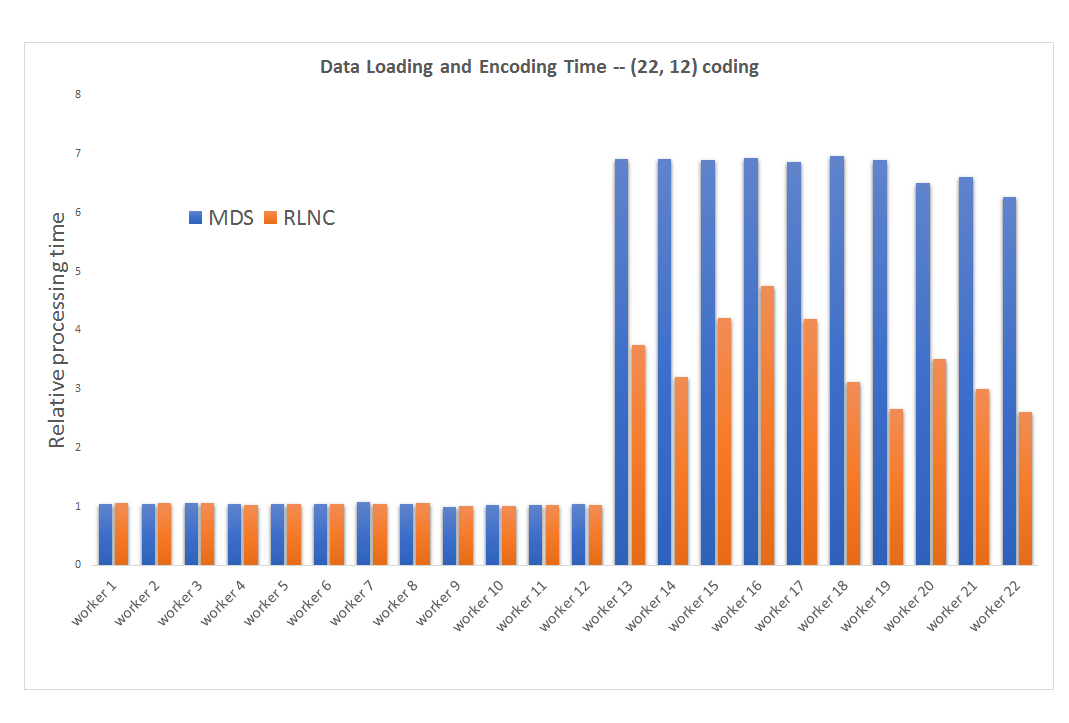}
\caption{Data loading and encoding time on each worker docker container for $(22,12)$ coding. On average, our RLNC can reduce the data loading and encoding time by \textasciitilde 50\%}
\label{encoding_22_12}
\end{figure}

Next we analyze the data loading and encoding time on each of Raspberry Pi nodes. As we discuss in the earlier section, mobile devices have  constrained resources. When compared to a server used in datacenter setting, the time for a Raspberry Pi node to load one partition of a matrix with $14,000$ rows and $5000$ columns is two orders of magnitude slower. And the latency will be even higher to linearly combine all $K$ sub-matrices and create an encoded sub-matrix at each of the $N-K$ redundant workers, if MDS is used. By using RLNC we are able to reduce data loading and encoding time by $50\%$ on average. The normalized time to load and encode data on all $N$ workers for both MDS and RLNC are shown in figure \ref{encoding_22_16} for the $(22,16)$ code used in logistic regression and figure \ref{encoding_22_12} for the $(22,12)$ code used in SVM. For the first $K$ workers, both coding strategies spend the same time for data loading because no encoding is required for these workers. For the remaining $N-K$ workers, RLNC reduces the data encoding time by $50\%$ on average. The variation in the data loading and encoding times seen in some of the worker nodes is a natural variation observed in our experimental setup due to various OS related events, such as docker management, garbage collection processes initiated by the OS and so on. 

\subsection{Logistic Regression results}
\begin{figure}
\centering
\includegraphics[width=.5\textwidth]{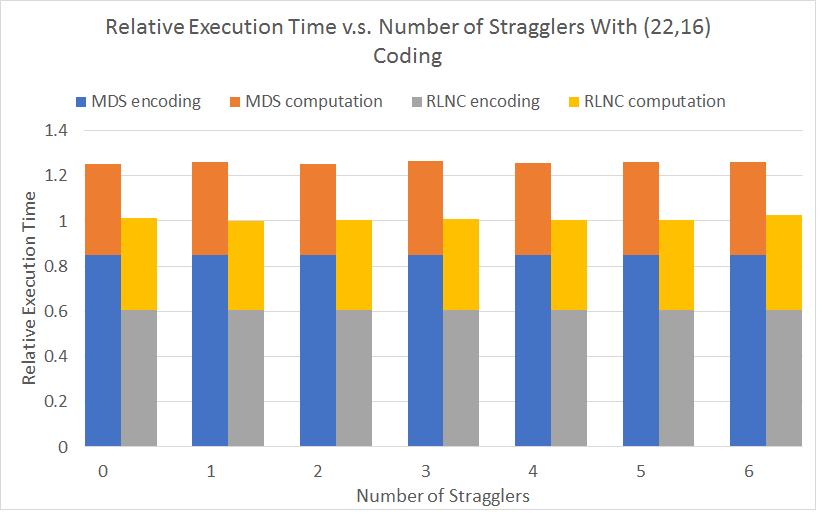}
\caption{Total execution time for Logistic Regression of $(22,16)$-RLNC v.s. $(22,16)$-MDS}
\label{fig:LR}
\end{figure}

In this section we analyze the time to execute 100 iterations of gradient descent for Logistic Regression on the 22-node Raspberry Pi cluster while using $(22,16)$ coding configuration. In figure \ref{fig:LR} we plot the relative execution time under $(22,16)$-MDS coding and $(22,16)$-RLNC  strategies while the number of stragglers are increased from 0 to 6. The execution time is the sum of the time spent on encoding and the time spent on computation i.e., performing 100 iterations of gradient descent. We can observe that the execution time of RLNC coded computation is \textasciitilde20\% lower than the corresponding MDS coded computation across all number of stragglers. This is due to the 50\% reduction in encoding time while using RLNC coding. And this is achieved without sacrificing any of the straggler tolerance capability in our experiments. This is because for RLNC on average, it only requires results from $0.032$ additional workers on top of $K=16$ workers for $(22,16)$-RLNC codes, which is a negligible overhead in practice. The actual matrix-vector multiplication time and vector decoding time are similar in both MDS and RLNC coding strategies.

\subsection{SVM results}
\begin{figure}
\centering
\includegraphics[width=.5\textwidth]{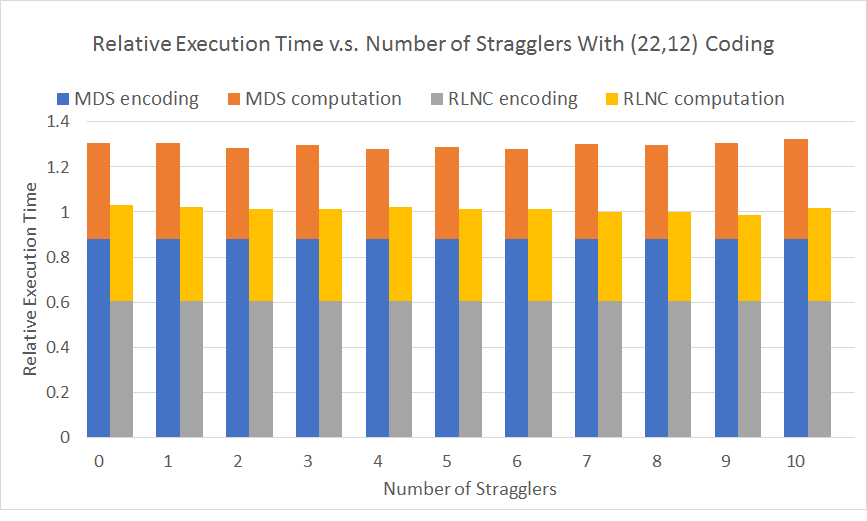}
\caption{Total execution time for SVM of $(22,12)$-RLNC v.s. $(22,12)$-MDS}
\label{fig:SVM}
\end{figure}

In this section we analyze the time to execute 100 iterations of gradient descent for Support Vector Machine on the 22-node Raspberry Pi cluster when using (22,12) coding configuration. In figure \ref{fig:SVM} we plot the relative execution time under $(22,12)$-MDS coding and $(22,12)$-RLNC strategies while the number of stragglers are increased from 0 to 10. The execution time includes the time spent on encoding and time spent on computation. We can observe that the execution time of RLNC coded computation is \textasciitilde25\% lower than the corresponding MDS coded computation across all number of stragglers. This is due to 50\% reduction in bandwidth (and the corresponding need to encode more data) time while using RLNC coding.  In our experimental setup RLNC  only requires  data from $0.2132$ additional workers while running SVM on top of $K=12$ workers for $(22,12)$-RLNC codes.   

\subsection{Training at Larger Scale}

\begin{figure}
 \centering
 \includegraphics[width=.5\textwidth]{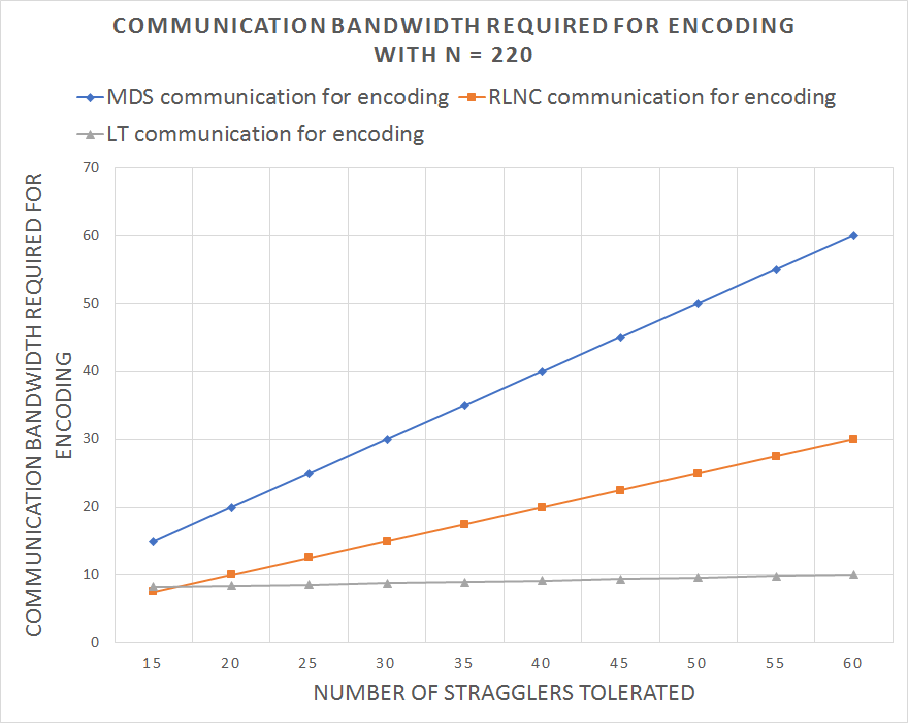}
 \caption{Communication bandwidth required for distributed encoding when we scale up to 220 nodes}
 \label{scale}
 \end{figure}

With the Kubernetes setup, we can expand our Raspberry Pi cluster systematically to hundreds of nodes and deploy much larger computing applications on them easily. As discussed in Section 4, the communication cost of MDS code and RLNC code is $K$ and $K/2$ per redundant worker. As such, the communication cost will increase linearly with $K$, which could be a concern in large scale deployments. This cost could potentially be reduced to $\log(K)$ per worker by using another coding technique called Luby-Transform (LT) code \cite{mackay2005fountain}, at a price of degraded straggler tolerance and additional encoding at the first $K$ workers. Although we don't have infrastructure to fully experiment with hundreds of nodes, we analyzed the communication bandwidth requirements in a scenario where we have 220 workers and the original data is distributed among 160 of these workers, meaning $N = 220, K = 160$, which is 10 times the scale of our Raspberry Pi cluster. Figure~\ref{scale} shows the bandwidth consumption can be significantly curtailed with LT codes when the system scales to even thousands of nodes with tens of straggles. 

\section{Related Work}
\noindent \textbf{Edge Computing}: Centralized cloud systems such as Amazon AWS and Microsoft Azure have become the de-facto platforms for data-intensive computing today. However, they suffer from inefficient data mobility due to the centralization of resources, which might prevent real time analysis because there is data transmission latency to and from the cloud. Authors in \cite{edgeCloudHirarchy} propose to deploy cloud servers at the network edge and design the edge cloud as a hierarchical tree of geo-distributed servers, so as to efficiently utilize the cloud resources to serve the peak loads from mobile users. On the other hand, decentralized mobile computing frameworks have been proposed to utilize mobile devices to solve dispersed-data-intensive applications, where the data may be spread at multiple geographical locations. Authors in \cite{Nebula} propose a dispersed cloud infrastructure that uses voluntary mobile computing resources for both computation and data storage. A similar model has also been presented in \cite{edgeCloud}, which aims to address edge computing specific issues by augmenting the traditional data center cloud model with service nodes placed at the network edges.

\noindent \textbf{Straggler Mitigation}: Straggler mitigation in distributed computing has received considerable attention and many techniques have been proposed in the literature. Effects of stragglers, at scale, like tail latency has been first demonstrated by authors in~\cite{tail}. They propose several software techniques to mitigate the effect of stragglers, such as redundant tasks and selective replication. The authors in the work~\cite{ananthanarayanan2010reining} studied the causes for stragglers like shared resource contention, disk failures etc., and propose schemes to quickly detect and cancel the stragglers. Replication of tasks to reduce the response times is studied in works such as \cite{clones,nihar,wang,gardner,chaubey,lee}. In this strategy, multiple replicas of each task are executed across worker nodes and results from the fastest copy are used. The slower replicas are killed.

\noindent \textbf{Coded Computation}: Coded computation leverages ideas from coding and information theory to improve the run-time performance of large-scale distributed computing. In particular, there have been two coded computing concepts proposed to deal with the communication and straggler bottlenecks in distributed computing. The type of coded computing proposed in ~\cite{LMA_all,li2016fundamental} trades off between computation load and communication load in distributed computing. This can be leveraged to speed up large-scale data analytics applications~\cite{CTS16}. Another type of coded computing ~\cite{speedUpML} provides resiliency in the presence of stragglers and can be utilized to mitigate tail latency ~\cite{speedUpML,reisizadehmobarakeh2017coded,LMA16_unify,dutta2016short,tandon2016gradient,polyCodes}. Authors in ~\cite{dutta2016short} propose a technique called short dot that allows individual worker nodes to perform a large number of the more efficient short products while only requiring a subset of them at the fusion node to decode/recover the final long dot product, while slightly compromising straggler tolerance. Authors of ~\cite{tandon2016gradient} propose and evaluate gradient coding schemes to provide tolerance against the presence of stragglers in distributed learning. Authors of ~\cite{reisizadehmobarakeh2017coded} proposes a coding framework for speeding up distributed computing with heterogeneous clusters in the cloud containing straggling servers. In this paper, we focus on the second type of coded computation (i.e., leveraging coding to mitigate straggler nodes) to provide robustness in mobile training. We propose, implement, and evaluate a coded computation framework to efficiently perform SVM and linear regression training using distributed mobile computing.

\section{Conclusion}
In this paper we built a scalable mobile device cluster consisting of Raspberry Pi 3 nodes and used it to analyze the challenges in machine learning training using mobile devices. We observed that the master node in the cluster becomes a bottleneck for encoding and explored distributed encoding strategies using MDS and RLNC codes. We demonstrated, through our experiments using gradient descent for Logistic Regression and SVM, that RLNC based distributed coded computation is upto 50\% more bandwidth efficient and provides upto 25\% better performance for gradient descent than MDS coded computation. In future work, we plan to scale out our edge cluster and expand our coded computing framework for more applications.

\bibliographystyle{sysml2019}
\bibliography{mobisys2018} 

\end{document}